On Site Raman analysis of rare Iznik potteries (late 15-16$^{th}$ centuries) as well as laboratory shard analyses show the stability of the glaze composition and the pigment originality. At least two different red glazes and two different green pigments are evidenced. Except for the period ~1510-1530, cassiterite was not used as an opacifier but to lighten the shade.

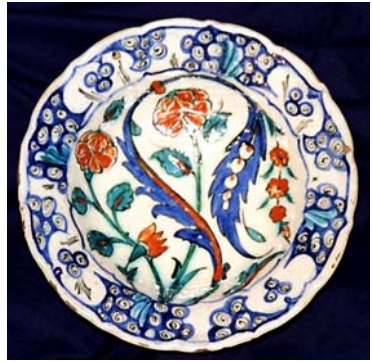

Ph. Colomban, V. Milande and L. Le Bihan

*On Site Raman Analysis of Iznik pottery glazes and pigments*

# On Site Raman Analysis of Iznik Pottery Glazes and Pigments

Philippe Colomban[a*], Véronique Milande[b] and Lionel Le Bihan[c]

[a]Laboratoire Dynamique, Interactions et Réactivité (LADIR)
UMR 7075 CNRS-Université Pierre et Marie Curie, 2 rue Henry Dunant, 94320 Thiais

[b]Musée national de Céramique
Place de la Manufacture, 92310, Sèvres

[c]Jobin-Yvon S.A.S
231 rue de Lille, 59650 Villeneuve d'Ascq, France.

**Abstract**

On site Raman analyses were performed at the Musée national de Céramique, Sèvres, France, on rare Iznik (former Nicaea) pottery produced from ~1480 to ~1620. A comparison is made with a series of shards. The town of production of these potteries was highly disputed in the 80's and many questions still remain. The potential of glaze on-site analyses as a classification/datating tool is evaluated. The structure of the silicate glaze does not change with the sample (index of polymerisation ~ 0.5-0.8, indicating a lead silicate composition; characteristic Si-O stretching mode doublet at ~985 and 1030-1050 $cm^{-1)}$. By contrast the corresponding signature of most of the "Kütahya" wares peaks at ~1070-1090 $cm^{-1}$. The lowest index is measured for a brilliant overglazed red bole, according to a lower temperature of (post)firing. The different crystalline phases identified in the glaze are α-quartz, haematite, spinel, cassiterite, uvarovite garnet and zircon. White colour arises from α-quartz slip in most samples studied. Cassiterite ($SnO_2$) opacifier is only present in some early blue-and-white ceramics (Master of the Knots and Baba Nakkas style, ca. 1510-1530) and we do not have other evidence of its intentional use as an opacifier. Intentional addition of tin oxide is likely for colour lightening in some red, blue and in clear green boles. At least two types of red glazes and two types of Cr-containing green pigments are evidenced.

*Corresponding author*
fax 33 1 4978 1118    colomban@glvt-cnrs.fr

## 1. INTRODUCTION

White, translucent, high temperature-fired ceramics were very rarely imported from China to Western countries before the Ming Dynasty [1-3]. They then became highly sought by Islamic Sultans and European Kings. First attempts to make equivalent porcelain were reported in Italy (Medici porcelains from ~1575 to 1587 [2,4]) but Iznik Ottomans had already been producing famous artefacts [1,3,5,6] in the style of Chinese and Vietnamese production for a century. Analyses of Iznik fritwares are rare (the main works were conducted in the 60's by Kiefer [7-9], Henderson [10] and Tite [11] in the 90's) and mainly focused on the body composition and microstructure. The origin of the "Iznik" production was highly disputed up to the 80's: production in Damascus (decorated with florist's flowers), Sivas, Iznik, Istanbul and Kütahya was reported; even Rhodes (artefacts with a famous red) was proposed in 19[th] century. Excavated ceramics in Iznik help to differentiate (former) Iznik from (later) Kütahya productions [6]. Recently, we demonstrated the potential of Raman (micro-)spectrometry for the classification of ceramics and glasses [12-15] and for a better understanding of the technology used to manufacture ancient objects [5,16]. Due to the very high value of Iznik potteries, the objects had to be kept in a secure area and it was not possible to take them to the LADIR for examination. We decided to attempt on-site examination using a portable instrument equipped with remote microscope objective. As a preliminary step, we selected artefacts produced during the late 15[th] and 16[th] –early 17[th] centuries, when the Iznik production was at its peak in quality. We simultaneously analysed a series of shards from the same period (except for a tile assigned to late productions) in the laboratory. Our purpose was to answer the following questions : i) what is the Raman signature of Iznik production ?, ii) was cassiterite ($SnO_2$) used as an opacifier, iii) what types of pigment were used ?, iv) Is it possible to definite criteria for a systematic study of Iznik production in order to propose a non-destructive procedure for classification and dating ?

## 2. EXPERIMENTAL
### 2.1. Samples

The objects analysed are presented on Plate 1. They belong to the collection of the Musée national de Céramique, Sèvres (France). The diameters range from ~ 15 cm (MNC 3840) to ~ 35 cm (MNC 3855). Among the samples studied is the famous Pilgrim's flask, inspired from Ming dynasty productions, made during the reign of Sultan Mehmed the Conqueror (ca. 1480-1490, published in ref. [6]), a blue and white dish (Master of the Knots style) ca. 1510, a nice blue and white vase inspired by 15[th] century Chinese/Vietnamese decor, ca. 1575 [17-20], dishes from ca. 1545 to 1580 with typical multicolour decor (flowers and leaves), a bowl (MNC 3840) and a jug decorated with typical green glaze and brilliant red bole overglaze (ca. 1580-90). The last dishes from ca. > 1600-1625 show a human figure (MNC 19565) and a lion (MNC 19577) depiction inspired from miniature paintings.

For comparison, we analysed at the laboratory a series of Iznik shards (Plate 2). The series contain a very characteristic red slip glazed shard ("c"), two dish pieces, the first ("a"), blue and white, in the Baba Nakkas style, typical of the potteries produced under Sultan Selim, ca. 1510-30, the second ("b"), polychrome, in the Potters' style, ca. 1540, with characteristic black overglazed lines and two tiles, one with flower décor ("d"), in which the transparent glaze covers all the decor, and the other ("e") with torsade, with black to green overglazed lines. Note, the white deposit only covers the area under the blue decor of shard "a".

### 2.2. Techniques

On-site micro-Raman spectrometry was performed using a transportable HE532 Jobin Yvon Raman system (Longjumeau, France). The analyser is optimised to work with a 532 nm laser excitation. It incorporates a unique concave holographic grating spectrograph. The optical construction is extremely stable and robust (no moving parts), with only two optical components within the spectrograph module (high luminosity). The specially developed concave holographic gratings provide unsurpassed sensitivity and total spectral coverage over the common Raman range (~160-3200 cm$^{-1}$). Peak intensity is however lowered by filtre absorption for wavenumber < ~200 cm$^{-1}$. The detector is a CCD multichannel matrix detector (1024x256 pixels) operating at -70°C (Peltier effect and air cooling). The advantage of remote measurements through optical fibres is well established and has proven its capability, e.g. in the study of Medici porcelain [5]. The HE instrument obtains spectral information from the sample via a single fibre optic probe such as the Jobin-Yvon SuperHead$^{TM}$ (weight ~500 g). This optic probe is a high efficiency Raman sensor, which enables *in situ*, non-invasive, chemical analysis. It is connected to the base analyser by fibre optic cables. The laser is also connected to this probe by a fibre. The probe contains an interference filter to remove the Raman signal from the fibre and the plasma lines from the laser. A notch filter is used to inject the laser inside the focusing objective and to transmit the Raman signal, which is collected by the objective and sent through a second fibre into the spectrograph. An Olympus MSPLAN$^{TM}$ Ultra Long Working distance objective (x80) was used. We used a portable XYZ plate (weight ~3 kg) to support the SuperHead allowing micronmetre scale displacements (see in Plate 1, k & k') to find a "good" location for spectral recording. Additionally the object itself was put on a support to be moved in different directions.

Typically, the power of illumination at the sample was around 20 mW. A density filter drastically decreasing the laser intensity was available to choose the area to be analysed and to focus the spot. Moving the SuperHead using the XYZ micrometer makes the optimisation. LabSpec® (Jobin-Yvon) software was used to control the HE system and the data acquisition. The total weight of the instrument was around 40 kg and wheels allowed easy transportation in the museum rooms.

Organic residues often cover the surfaces of ancient ceramics. Some organic groups are grafted to the pore surface and promote fluorescence, which sometimes hinders the recording of the Raman spectrum. Yet the rather large power of illumination (~20 mW) through a microscope objective cleans the surface very well. In many cases the use of high magnification objectives (here x80) also helped to find locations free of any fluorescence. This was successful for the examination of Iznik glaze. By contrast, strong fluorescence prevented the analysis of the body phases. Note body analysis has been successfully done on Medici porcelain using red excitation [5]. Strong background-free spectra are obtained when the laser spot is focused on strongly coloured decor: 5-20 accumulations of 5-50 s were typically used to record the spectra (see for instance the upper spectra of Figures 1-4). On the contrary the collection of the colourless glaze signature is more difficult and requires a larger number of accumulations (>100) with shorter recording time to avoid saturation (higher fluorescence background).

**2.3. Peak fitting and data processing**

In undertaking a curve fit of the Raman spectra, the linear baseline was first subtracted using LabSpec® software. The same spectral windows were used for the extraction of the components using the Origin® software peak-fitting module (Microcal Software, Inc.). The integral area, the bandwidth and the centre of gravity were calculated for each component. A Gaussian shape was assumed for all Raman lines because of the amorphous state of examined materials (a Lorentzian shape is used for crystalline phase components). The following assumptions were made: i) for the Si-O stretching range extending from 700 to 1300 cm$^{-1}$, (see further) we postulated 5 components assigned to $Q^0$, $Q^1$, $Q^2$, $Q^3$ and $Q^4$ according to the number of oxygen atoms bonded per tetrahedron, as described in previous papers [12,21,22] ii) we postulated rather similar bandwidths for the Si-O bending range, the number of bands also being 5; one or two additional narrow bands were added, if necessary, to take crystalline phases ($\alpha$–quartz, peak at 464 cm$^{-1}$) and pigment signatures (cassiterite, spinels, …) into account.

## 3. RESULTS AND DISCUSSION
### 3.1. Colourless/ white Glaze

Analysis of the colourless/white regions of the samples always reveals the more or less strong Raman signature of $\alpha$-quartz (Figs 1-5), with the typical peaks at 205, 265, 355, 400, 465, 695, 805, 1085 and 1160 cm$^{-1}$ [23-26]. By modifying the focal point of the laser with the XYZ plate it is possible to confirm that the white colour is obtained by an $\alpha$-quartz-rich slip deposited on the pink-to-yellow body, below the glaze, as reported in previous studies [6-10]. Note, the technique of a white $\alpha$-quartz slip to mask the red colour of the body was already used for Vietnamese porcelain and stoneware [27]. The Raman signature of cassiterite (precipitate of $SnO_2$ in the glaze) is only observed from place to place in shard "a" (white deposit below the blue decor) and in MNC 5172 dish, with the narrow doublet at ~635 (strong) and 775 (weak) cm$^{-1}$ [15]. A rather narrow peak is also observed at ~650 cm$^{-1}$ in the white glaze of the bowl (MNC 3840). Assignment to cassiterite is possible, but the wavenumber shift would indicate a tin oxide-based solid solution (with addition of antimony oxide (?) [15]). Synthesis of reference materials is now mandatory for comparison and definitive assignment.

The very heterogeneous distribution of cassiterite in shard "a" and in MNC 5172, at the (sub-)millimetre scale, indicates that cassiterite was not obtained by homogeneous nucleation on cooling of the glaze, but results from the primary deposit of an $SnO_2$-rich mixture before glaze enameling.

Considering the Si-O bending (~500 cm$^{-1}$) and stretching (~1000 cm$^{-1}$) envelopes [12,21,22], glaze signatures are rather homogeneous, both for artifacts (Figs 1-4) and shards (Figs 5-8). All samples have their main components at 1035-1055 cm$^{-1}$ (Table 1), except the shard "e" (torsade), which shows a strong 1095 cm$^{-1}$ component. This last spectrum is very similar to those recorded on many Kütahya wares. It is obvious that this sample has been prepared with a different technology. Two-peak Si-O stretching fingerprints are usually observed in mixed lead-alkaline glasses [7-10]

### 3.2. Blue glaze and pigments

Strongly coloured regions give a strong Raman signature, nearly free of any background. The ways to obtain a blue glaze are rather limited: copper in alkaline silicate (turquoise), cobalt [15,28-30] (blue) and lapis lazuli [16,31,32] (ultramarine blue), whatever the fluxing agent. The last pigment has a very typical resonance Raman signature [16]. If small amounts of cobalt (copper) dissolve in the glaze, no specific Raman signature is expected. Saturation leads to cobalt mixed silicates (olivine, main peak expected at ~ 820-880 cm$^{-1}$ [24]) or to cobalt aluminate (main peak at 208 and 525 cm$^{-1}$). Cobalt-containing Mn-rich ores (spinels) can also be used as blue pigments [19], if the firing is conducted in a reducing atmosphere (to avoid colouration by $Mn^{n+}$ ions). The observed colour excludes the use of lapis lazuli and Cu. In most blue regions analyzed, we only observed the Raman signature of the glaze: this indicates that cobalt is dissolved within the glaze network. Note that the centre of gravity of the stronger Si-O stretching component peaks at ~1065-1070 cm$^{-1}$ (Fig. 1) for pigment-rich and quartz-rich regions, instead of ~1055 cm$^{-1}$ for colourless glaze. This indicates that the composition of the blue glass used as pigment differs from that of the overglaze. We also noted a small peak at ~645-650 cm$^{-1}$ in the Pilgrim's flask glaze (Fig. 1). As proposed above, this peak could correspond to traces of cassiterite solid solution ($SnO_2$–$Sb_2O_3$ ? [15]). A clear signature of cassiterite is only observed in blue design of shard "a" (Baba Nakkas style ware) and in the MNC 5172 dish from nearly the same period (ware of the Master of the Knots style). Note zircon traces ($ZrSiO_4$, mean peaks at 357 and 1007 cm$^{-1}$ [24]) are observed in some places in the blue decor in shard "a". Zircon traces are characteristic of some sand and clays. Traces of cassiterite and α-quartz are observed in the turquoise florist's flower of MNC 3855 dish (Fig. 2) and in the white glaze.

**3.3. Red glazes**

Red colour is famous in the Iznik palette and the colour shade is a criterion for dating [1,6]. The Raman signature of haematite (α-$Fe_2O_3$) is clearly observed (peaks at 220, 300, 410 and 1320 $cm^{-1}$ [15]), for instance in the jug (MNC 26230, Fig. 4) and shard "d" (Figure 8). Haematite is associated with one or several other phases for lightening the colour: cassiterite in the bowl MNC 3840 (Fig. 3), α-quartz in the jug MNC 26230 (Fig. 4) and shard "d" (Fig. 8). Extending the series of measurements could demonstrate if the relative amount of the associated phases is a criterion for dating.

**3.4. Green glaze and pigments**

Analysis of the green regions offers a great variety of spectra. The main peaks are observed at 635 or 640-695 $cm^{-1}$ (multicolour dish MNC 27397, Fig. 3), 700 $cm^{-1}$ (Fig. 2, MNC 38555 blue dish), 845-855 $cm^{-1}$ (lion dish MNC 19577, jug MNC 26230, Fig. 4, shard "d", Fig. 7). Different ways of obtaining the green colour are possible: dissolution of Cu in the lead-based glaze [15,28-30] (no specific Raman signature, e.g. Fig. 3, MNC 27397), yellow pigment (Naples yellow $Pb_2Sb_2O_7$, main signature at ~130-145 and 510 $cm^{-1}$ [15,23,33,34]) mixed with blue Co-coloured glaze [15], use of Cr-containing pigments such as Cr-doped wollastonite (main narrow peaks at 578 and 985 $cm^{-1}$ [14]), uvarovite garnet (also called Victoria green pigments, main peaks at 820-850 $cm^{-1}$ [10]) and spinels (main peak at ~680-700 $cm^{-1}$) or corundum (e.g. $Cr_2O_3$, main peak at 613 $cm^{-1}$) [15,35,36]. It is clear that a careful study of the green glaze will offer a way of classifying the samples. Note that cassiterite is evident in the light green regions of the MNC 27397 dish (Fig. 3 & Table 1). The use of $SnO_2$ to enlighten the shade is thus obvious.

**3.5. Other colours**

Dark pink (Fig. 2, MNC 3855 dish, we did not succeed in getting a specific signature of this colour) and black (Figs. 2, MNC 3855; 3, MNC 27397; 4, MNC 26230) are also observed. The spectra of black lines and dots are similar to those of the green decors. This is consistent with the proposition from Kiefer [37] that chromite, a Cr,Al,Fe containing-pigment, was used. Additionally, traces of carbon are observed in MNC 27397 dish (doublet at 1360 and 1600 $cm^{-1}$, Fig. 3). Is this carbon trace a proof of *cuerda seca* technique (use of a volatile element to separate the coloured area or addition of an organic material to prevent from the oxidation) ?

**3.6. Comparison with previous analyses**

Kiefer [7-9] and Hendersen [10] measured the composition of body and glazes of shards assigned to Iznik productions. Typically the mean composition of Iznik glaze is: $SiO_2$ 45-47 wt%, $Na_2O$ 8-14 wt% ($K_2O$~1wt%, CaO~1wt%) PbO 25-30 wt%; $SnO_2$ content varies between 4 and 7 wt%. The index of polarization of the glaze, measured from the area ratio of Si-O bending to stretching envelopes [21], ranges between 0.4 and 0.8 (Fig. 9). From previous established relationship between the index and the glazing temperature [21], a glazing temperature close to 700-800°C, or less can be proposed. Note that larger values are measured for the pigment-rich (or eavy coloured) and quartz-rich glazes. Lowest values are measured for the red bole, indicating a very low firing temperature. The comparison of the Raman spectra shows a good homogeneity of the glaze structure. However, by plotting $Q_1$ centre of gravity as a function of $Q_2$, one clearly puts shard "a" apart (Fig. 9).

Analyzing the glaze from the top to the α-quartz slip shows a shift of the Si-O stretching wavenumber, from ~1040-1055 $cm^{-1}$ to 1075 $cm^{-1}$, typically. This indicates that the (over)glaze composition is different from that used to prepare the pigment or mixed with the slip. Different explanations can be proposed: i) the composition around pigment and quartz slip grains changes because a special mixture was applied, ii) the composition changes because the grains dissolve in the glaze flux, iii) some surface corrosion.

Because Sn was detected in elemental analysis, many authors [1,3,37-39] concluded that cassiterite ($SnO_2$) has been used as opacifier, although the white α–quartz slip is also well documented [1,6]. Except in some early artifacts (shard "a" and MNC 5172 dish) it is clear that cassiterite is either absent from, or present as traces only, in the white glaze of our samples. We can conclude that tin is an impurity in the lead providing ore (the use of bronze slag, the so-called *calcine*). Note that recent combined EDX and Raman analysis of polychrome (blue and green touchs on cream) earthenwares (copies of Tang ceramics made in Mesopotamia, 9[th] century), that are believed to be the first "faience" (Sn was detected in the glaze) demonstrated that Sn is not present in the form of a precipitate of cassiterite (making the glaze opaque and white) but as a mixed Ca-K-Sn compound [40]. On the contrary it is clear that some red and green boles contain cassiterite and this could be an intentional use in order to lighten the shade. A systematic study of the red decor is mandatory to give a definite conclusion.

Henderson observed the element chromium in the green decor of one sample (dated 1650). It is not clear whether this element has been searched for systematically in all the samples studied of the Henderson series [37]. At least two types of pigment were used to obtain the green decor of the samples studied in this work: i) a spinel (MNC 8408, MNC 3855) and ii) a pigment of uvarovite garnet-type (MNC 27397, MNC 26230). All Raman signatures confirm that phases with the element chromium are used. EDS analyses, are however, required for a definitive assignment. The crossover between the two cases seems to correspond to date ~1550. A systematic study and a comparison with synthetic pigment, well characterized using both X-ray and Raman techniques would be useful to progress in the technology understanding. Note that the black lines were obtained in the same way by use of nearly pure pigment (Fig. 3, MNC 27397).

Microprobe quantitative analysis of the black pigment used in the 17[th] Safavid blue and white ceramics has been published recently [41]. Very heterogeneous compositions were found but they are all compatible with the chromite spinel structure : $Cr_2O_3$ 46-60%, $Fe_2O_3$ 6-14%, CoO 0-17%, MgO 7-18%, $SiO_2$ 0.1-8%. Two sorts of geological chromites ($MgCr_2O_4$ and $FeCr_2O_4$ end members) are usually found, in Persia for instance. It is obvious that a selection of ore could give different colours in the glaze, from blue-green (Mg/Co-rich) to black (Fe-rich). Similar conclusion stands for Iznik pottery. Like the Persian potter, the Iznik potter used black lines (which do not alter on firing) to highlight designs (see Plate 1 d-g and plate 2 b-e) and to avoid any diffusion of the colour. In this way the technique plays the same role that the *cuerda seca* work of some tiles and majolicas. On the contrary, it is obvious from the sample examination that blue (and green) colour diffuse in the glaze, especially in the later productions (Plate 1k & k'). The chemical explanation is simple: Co (and Cu) are present as isolated ions and diffuse in the glaze easily and the sharpness of the drawing is lost. By contrast chromite spinel is a stable pigment; its high melting temperature prevents strong reaction with the molten glaze on firing. Furthermore $Co^{2+}$ ions could be fixed in the chromite phase. For instance the black line of figure 3 (MNC 27397) gives a spectrum of the pure pigment (strong single 845 cm$^{-1}$ peak) whereas the dark-green region gives the spectrum of a glaze-quartz-pigment mixture. Peaks in the 820-860 cm$^{-1}$ range are characteristic of $CrO_4^{2-}$ ion-containing structures [42], but center of gravity, intensity and splitting show subtle difference as a function of exact composition and structure. It is obvious from this preliminary Raman report that different pigments are used to give the green colour. Two explanations are possible: i) the use of specific ore for each colour, ii) a special preparation of the ore or a true synthesis of pigment. The strong similarity between the observed spectra and those recorded on Victoria pigments (Cr-doped calcium silicate garnet) and corresponding glaze prepared at the Manufacture Nationale de Sèvres [15] makes us believe that a specific pigment was prepared. The synthesis of reference pigments for Raman analysis is now required to progress the sample analysis.

Different types of red were evidenced as a function of haematite/quartz/cassiterite relative proportion: i) a red glaze by mixing α-quartz with a low proportion of haematite and glaze matrix (shard "d"), ii) a darker glaze containing a lower quantity of α-quartz (MNC 2630), iii) a very special red (MNC 3840) made by mixing α-quartz + haematite with a modified cassiterite-containing glaze. Raman measurement of the relative phase proportions is in progress.

**4. CONCLUSION**

This preliminary on-site study of a small selection of ceramics assigned to Iznik production confirms the potential of the Raman technique as a non-destructive tool for classification (and, possibly for dating). Except for shard "e" (strong peak at 1095 cm$^{-1}$), the Raman signature of the glaze remains constant (characteristic broad doublet at ~985 and 1040-1060 cm$^{-1}$), according to a lead-rich composition, rather similar to that used in soft-paste French porcelain [14]. The signature of shard "e" is very similar to those measured for most of the "Kütahya" ware [43] and can be compared to that of mixed silica-rich (lead-)alkaline glazes [13,14]. This deserves further study of representative Kütahya productions. Examination of the Iznik glazes shows a clear increase of the gloss quality with times, indicating better firing control then declines in quality in the first decades of the 17$^{th}$ century. On the contrary, large changes in pigment technology are observed. At least two different red glazes and two different green pigments are evident. Except for the period ~1510-1530 cassiterite was not used as an opacifier. A systematic study of cassiterite precipitate, red and green decors as well as a comparison with synthetic pigments is in progress to go further into the understanding of the ancient technology of Ottoman potters.

**ACKNOWLEDGMENTS**

The authors thank the Museum Director, Mrs Antoinette Hallé for her help and support. Special thanks to Mrs. Marthe Bernus-Taylor, former Head of the Louvre Museum Islam Section for many discussions and to Mrs Laure Soustiel and Marie-Christine David, Islamic Art Consultants, for the provided shards and help in dating of the artefacts.

**REFERENCES**

1. Soustiel J. *La céramique islamique – Le guide du connaisseur*, Office du Livre – Editions Vilo, Paris, 1985.
2. Lane A, *Italian Porcelain*, Faber and Faber, London, 1974.
3. Caiger-Smith A, *Tin Glaze Pottery in Europe and Islamic World*, London, Faber & Faber, 1973.
4. Kingery DW, Smith D, *The development of European soft-paste (Frit) porcelains*, in *"Ancient Technology to Modern Science"*, vol. I, Ceramics and Civilization, Kingery DW Ed, The American Ceramic Soc., Columbus, OH, pp273-292, 1984;
5. Colomban Ph, Milande V, Lucas H, *J. Raman Spectr*. 2004; **35**: 68.
6. Atasoy N, Raby J in *Iznik, The pottery of Ottoman Turkey*, Petsopoulos Y Ed., Alexandria Press, London, 1989.
7. Kiefer Ch, *Cahiers de la Céramique* 1956; **4**: 15.
8. Kiefer Ch, *Bull. Soc. Franç. Ceram.* 1956; **30**: 3.
9. Kiefer Ch, *Bull. Soc. Franç. Ceram.* 1956; **31**: 17.
10. Henderson J, ch VI, in ref [6], p. 65.
11. Tite MS, *Archaeometry* 1989; **31**: 115.
12. Colomban Ph, Treppoz F. *J. Raman Spectr*. 2001; **32**: 93.
13. Colomban Ph, March G, Mazerolles L, Karmous T, Ayed N, Ennabli A, Slim H. *J. Raman Spectrosc*., 2003; **34**: 205.
14. Colomban Ph, *Glasses, Glazes and Ceramics – Recognition of the Ancient Technology from the Raman Spectra*, in *Raman Spectroscopy in Archaeology and Art History*, Edwards HGM and Chalmers JM (Eds), Royal Society of Chemistry, London, 2004.
15. Colomban Ph, Sagon G, Faurel X. *J. Raman Spectr*. 2001; **32**: 351.
16. Ph. Colomban, *J. Raman Spectr*. 2003; **34**: 420.
17. Quingzheng W, *A dictionary of Chinese Ceramics*, Sun Tree Publishing Ltd, Singapore, 2002.
18. Tri B.M., Nguyen-Long K. *Vietnamese Blue and White Ceramics*. The Social Sciences Publishing House, Hanoi 2001.
19. Liem NQ, Colomban Ph, Sagon G, Tinh HX, Hoang TB, *J. Cult. Heritage* 2003; **4**: 187.
20. Stevenson J, Guy J, *Vietnamese Ceramics: a separate tradition*, Art Media Resources, Avery Press, Chicago, USA, 1997.
21. Colomban Ph. *J. Non-Crystalline Solids*, 2003; **322**: 180.
22. Liem NQ, Thanh NT, Colomban Ph, *J. Raman Spectrosc*. 2002; **33**: 287.
23. Bell IM, Clark RJH, Gibbs PJ, *Spectrochemica Acta,* 1997; **53**: 2159.
24. Griffith WP, ch 12. in *"Infrared and Raman Spectroscopy of Lunar and Terrestrial Minerals"*, Ed. Karr C Jr, Academic Press: New York, 1975.
25. Pinet M, Smith DC, Lasnier B. *La Microsonde Raman en Gemmologie, Revue de Gemmologie*. 1992 (Hors Série, Paris, June).
26. Maestrati R . *Diplôme d'Université de Gemmologie*, Université de Nantes, 1989.
27. Colomban Ph, in *Arts du Vietnam – La fleur du pêcher et l'oiseau d'azur*, Noppe C and Hubert J-F Eds, La Renaissance du Livre, Tournai, 2002 ; pp100-109.
28. Eppler RA, Eppler DR. *Glazes and Glass coatings*, The American Ceramic Society, Westerville, Ohio, 2000.
29. Deck Th, *La Faïence*, Maison Quantin, Paris, 1887.
30. Bertan H, *Nouveau Manuel Complet de la Peinture sur Verre, sur Porcelaine et sur Email*, Encyclopédie-Roret, Mulo L Ed., Paris, 1913.

31. Clark RJH, Curri ML, Laganara C, *Spectrochimica Acta A* 1997; **53**: 597.
32. Freestone IC, Stapleton CP, Composition and Technology of Islamic enameled glass of the 13-14$^{th}$ centuries (ch 24) in *Gilded and Enamelled Glass from the Middle East*, Ward R Ed, British Museum Press, London, 1998.
33. Colomban Ph, Sagon G, Louhichi A, Binous, H, Ayed N. *Revue Archéométrie* 2002; **25**: 101.
34. Sakellariou K, Miliani C, Morresi A, Ombelli M, *J. Raman Spectrosc.* 2003; **34**: .
35. Wang Z, O'Neill HSC, Lazor P, Saxena SK, *J. Phys. Chem. Solids* 2002; **63**: 2057.
36. Colomban Ph, Jullian S, Parlier M, Monge-Cadet P, *Aerospace Sci & Techn.* 1999; **3**: 447.
37. Kiefer Ch, in ref 1 p 368.
38. Cooper E, *Ten Thousand Years of Pottery*, University of Pennsylvania Press, Philadelphia, 4$^{th}$ edition, 2000.
39. Allan J.W., Islamic Ceramics, Ashmolean Museum Oxford, 1991.
40. Colomban Ph, Truong C, *J. Raman Spectrosc.* 2004; **35**: .
41. Agosti MD, Schweitzer F, in "Persia and China – Safavid Blue and White Ceramics in the V&A Museum, 1501-1738", Crowe Ed., Thames & Hudson Ltd, Farnborough, 2002.
42. Frost R.L., *J. Raman Spectrosc.* 2004; **35**: 153.
43. Ph. Colomban, V. Milande, unpublished work.

# FIGURES CAPTIONS

**Plate 1** : Master pieces from the Musée national de Céramique, Sèvres (Photo © Ph. Colomban)
a) Blue and white Pilgrim's flask, ca. 1480 (-1490 ?), Baba Nakkas-Rumi-Hatay style made under Sultan Mehmed the Conqueror, MNC 15472, Piet-Lataudrie bequest (1910); a') detail of the on-site examination. (diameter ~30 cm ).
b) Blue and white dish (diameter ~40 cm), Master of the Knots style, ca. 1510-1520; MNC 5172, Troloppi Coll., acquired 1858; b') back side view.
c) Blue and white vase, Chinese or Vietnamese copy style, ca. 1560-1580 (this shape was formed after ~1520), MNC 8408, Davillier bequest (1885); (h~38 cm).
d) Polychrome jug, ca. 1580-1590, decorated with red bole overglaze; MNC 26230, Mrs M. Herisson bequest (1988); (h~13 cm)
e) Large blue "Damascus" ware style dish decorated with white florist's flowers and softly foliated rim, ca. 1550-55, MNC 3855, Coll. Perret, acquired 1848; (diameter~35 cm); e') backside view.
f) Large dish, Damascus' ware style, ca. 1545-1550, decorated with green flower, Saz leaf and softly foliated rim, MNC 8411, (diameter: ~39 cm); f') backside view, Davillier bequest.
g) Bottom of a covered basin on hight foot, (Rhodian) floral style style, ca. 1560-80, MNC 3840, acquired 1848 (diameter : ~18 cm); g') inside view.
h) Polychrome dish decorated with flowers and Saz leafs, with a Chinese style wave-and-rock pattern border, ca. 1575, MNC 18853, from the Musée de Cluny collection (diameter : ~20 cm).
i) Polychrome dish decorated with red clove pink, ca. 1575-1580, MNC27397, from the Musée de Cluny collection; i') backside view (diameter : ~28.5 cm).
j) Polychrome dish decorated with human single-figure, ca. 1600-1625, MNC 19565, from the Musée de Cluny collection (diameter ~30 cm); j') Polychrome dish decorated with a lion, ca. 1600-1625, MNC 19577 (diameter ~30 cm).
k) and j'), example of on site analysis.

**Plate 2** : Shards (Private Coll.; photo © Ph. Colomban)
a) Blue and white bowl bottom, Baba Nakkas style ware, made under Sultan Selim, ca. 1510-30 (~10x8 cm$^2$; thickness 0.5 cm).
b) Polychrom dish rim, Potters' style, ca. 1535-45; b') backside view (~6x5 cm$^2$; t=0.5 cm). Note the black overglazed lines.
c) Red slip decorated dish bottom (R), ca. 1550-1560 (~7x7 cm$^2$; t= 0.7 cm). Note the blue and white overglazed decor.
**d)** Polychrom tile shard decorated with flower, ca. 1580-1600 (~8x7 cm$^2$; t=1.4 cm). Note the optically clear glaze cover all the surface.
**e)** Polychrom tile shard with torsade decor (T) > 1600 (~7x9 cm$^2$; t= 1.2 cm). Note the black and green overglazed lines.

**Fig. 1** : Representative Raman spectra recorded on blue and white glazed regions of the vase (MNC 8408) and Pilgrim's flask presented in Plate 1 (a & c). Except for the top spectrum, a linear base line has been subtracted. Top spectrum was recorded at the glaze surface, over the pigment-containing glaze.

**Fig. 2** : Representative Raman spectra recorded in various places of the "Damascus" blue dish verso (MNC 3855, Plate 1e). Except for the top spectrum, a linear base line has been subtracted.

**Fig. 3** : Representative Raman spectra recorded in various places of the bowl (MNC 3840, Plate 1g) and dish MNC 27397 (Plate 1i). A linear base-line has been subtracted for some spectra.
**Fig. 4** : Representative Raman spectra recorded in various places of the jug (MNC 26230, Plate 1d). Except for the top spectrum, a linear base-line has been subtracted.
**Fig. 5** :  Representative Raman spectra recorded on the "white" (red) glaze for the shards presented in Plate 2 (a, b, c and e). A linear base-line has been subtracted.
**Fig. 6** :  Representative Raman spectra recorded on blue regions of e shard a. A linear base-line has been subtracted.
**Fig. 7** : Representative Raman spectrum recorded in the green glaze of shard d. A linear base-line has been subtracted.
**Fig. 8** : Representative Raman spectrum recorded in the red glaze of shard d. A linear base line has been subtracted.
**Fig. 9** : Index of polymerization ( $A_{500}$ / $A_{1000}$) measured for the shard glaze.
**Fig. 10** : Bivariate plot of the Si-O stretching $Q_1$ and $Q_2$ components for the shard series (Table 1); shard a: aw, white region and ab, blue region; shard c: c, red region.; shard d: white region and dg, green region.

**Table 1 : Centres of gravity of the main components extracted from the Raman spectra of the shard and artifact glazes**

| Sample Colour+ | | | | Wavenumber / cm$^{-1}$ | | | | | | | |
|---|---|---|---|---|---|---|---|---|---|---|---|
| | | | Q'$_n$ | | | | Q$_0$ | Q$_1$ | Q$_2$ | Q$_3$ | Q$_4$ |
| a, white | 240 | 380 | 477 | 542 | 585 | | 780 | 940 | 990 | 1045 | 1107 |
| b, white | | 415 | 477 | 537 | 604 | | 783 | 923 | 980 | 1047 | 1113 |
| c, red | | | 398 | 470 | 525 | | 780 | 920 | 975 | 1035 | 1115 |
| e, white | | | 398 | 475 | 553,580 | 640 | 785 | | 998,1005 | 1095 | 1170 |
| a, blue | | | 390 | 482 | 545 | 595,**_635_** | 789 | 935 | 981 | 1043 | 1130 |
| d, green | 352 | | 392 | 476 | 545 | 620 | 795,**_847_** | 915 | 985 | 1045 | |
| d, red | | | 413* | | | 615 | 793 | 925 | 975 | 1045 | 1160 |
| 8408, blue1 | 227 | 412 | **_465_** | 523 | 635 | | 783 | 918 | 978 | 1055 | 1136 |
| " , blue2 | | | 483 | 512 | 550 | **_643_** | 789 | 924 | 985 | 1075 | 1164 |
| " , blue3 | 315 | | 466 | 510 | 550 | **_650_** | 788 | 924 | 983 | 1065 | 1164 |
| 3855, black 1 | | | 456 | 528 | 547 | 658 | 790 | 935 | 980 | 1041 | 1111 |
| " , blue1 | | 390 | 482 | 545 | 655 | | 786 | 926 | 978 | 1048 | 1140 |
| " , green1 | | | 483 | 550 | 650 | | 790 | 930 | 979 | 1041 | 1100 |
| 3840, yellow | | 446 | 466 | 529 | 635 | | 783 | 920 | 978 | 1055 | 1196 |
| " , red | | 426 | 466 | 528 | **_635_** | | 786 | 917 | 980 | 1053 | 1128 |
| 26230, blue1 | | 458 | **_465_** | 527 | 633 | | 782 | 921 | 976 | 1050 | 1149 |
| " , red1 | | 419 | **_465_** | 514 | 645 | | 791 | 926 | 975 | 1045 | 1154 |
| " , green1 | | | 466 | 502 | 548 | 648 | ,**_846_** | 920 | 980 | 1050 | 1160 |
| 27397, green1 | | 402 | 465 | 528 | **_635_** | | 779 | 915 | 970 | | 1113 |
| " , black1 | 351 | 401 | 509 | 562 | 650 | | 705, **_844_** | 900 | 930 | | |
| ", dark-green1 | | 382 | **_465_** | 502 | 645 | | ,**_845_** | 913 | 970 | 1036 | 1144 |

**_Number_** : narrow peaks of crystalline pigments (cassiterite, spinels, α-quartz, etc)
+ colour of the analyzed area

Table 2 : Main conclusions relevant to the history of ceramic technology

| Question | Evidence | Implication on technology / history |
|---|---|---|
| $SnO_2$ opacification | Strong cassiterite Raman signature | Only during ~ years 1510-30 Master of the Knots & Baba Nakhas styles |
| Intentional $SnO_2$ addition for Blue, green and red lightening | weak cassiterite signature | Advanced pigment technology |
| Use of different pigments for the same colour, but different shade | spinel and garnet signature haematite, α-quartz and glass | Advanced pigment technology |
| Differentiation between Iznik and Kütahya wares | Si-O stretching mode fingerprint | Origin |

a) MNC 15472 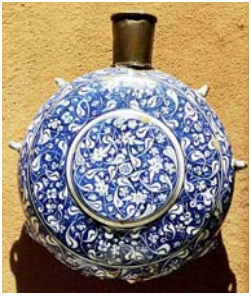 a') 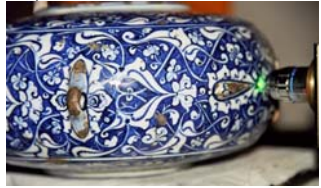

g) MNC 3840 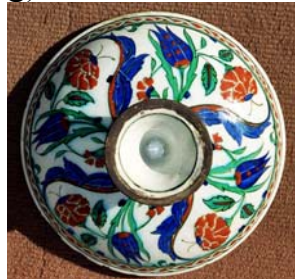 g') 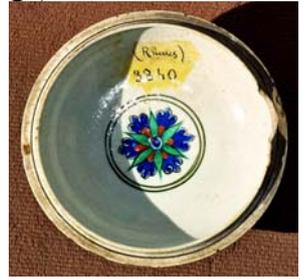

b) MNC 5172 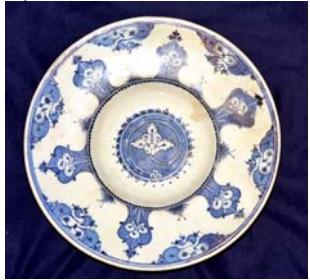 b') 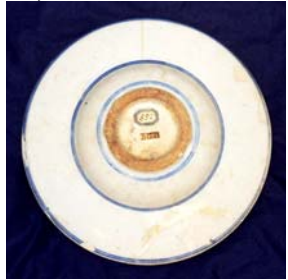

h) MNC 18853 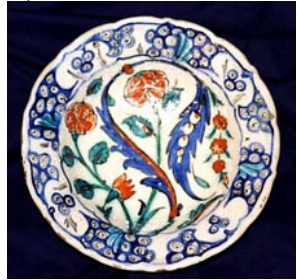 h') 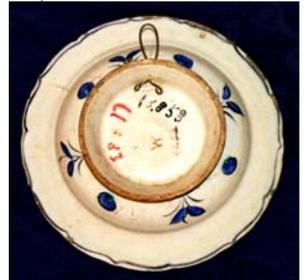

c) MNC 8408 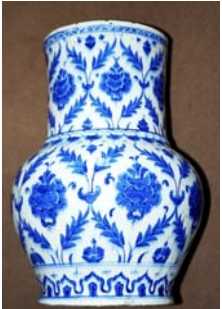

d) MNC 26230 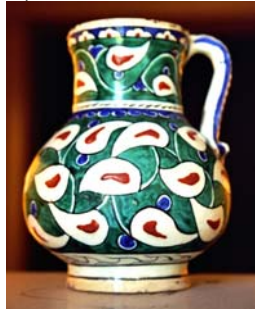

i) MNC 27397 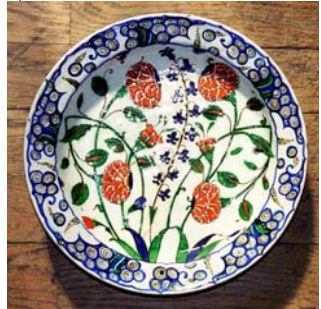 i') 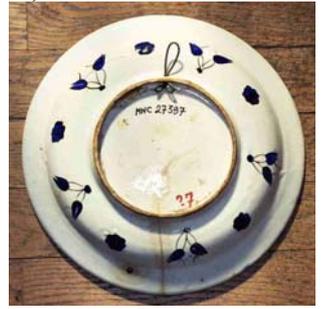

e) MNC 3855 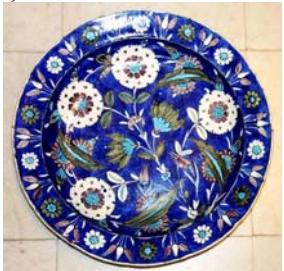 e') 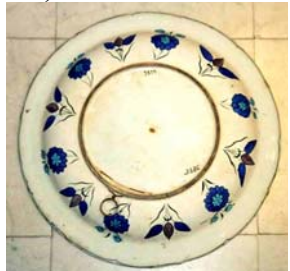

j) MNC 19565 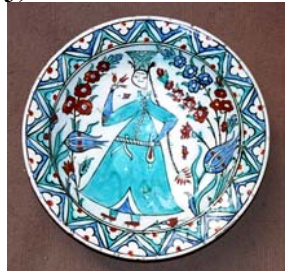 j') MNC 19577 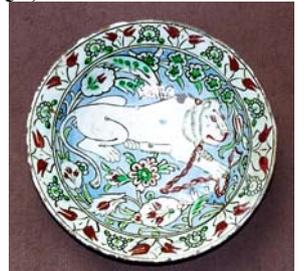

f) MNC 8411 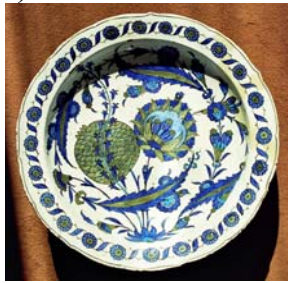 f') 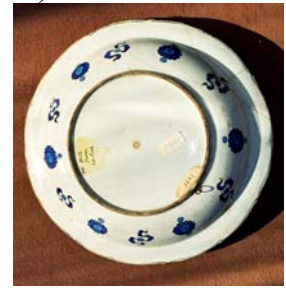

k) 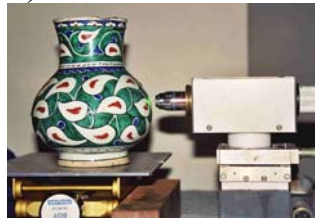 k') 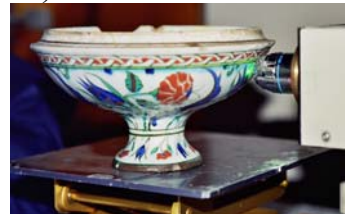

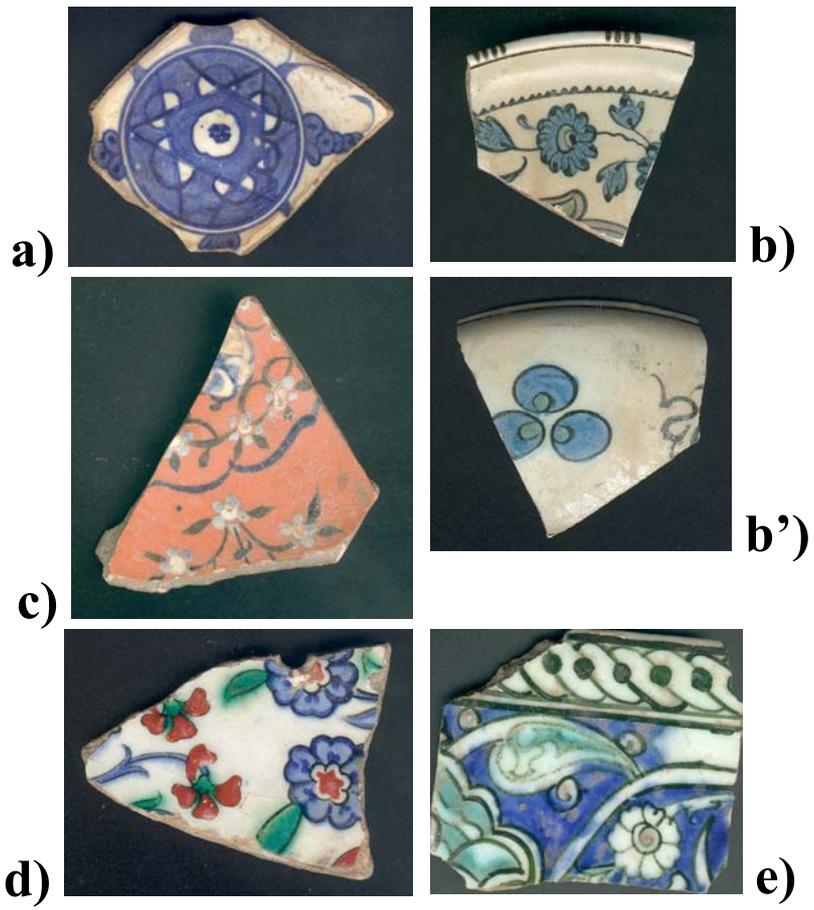

Plate 2

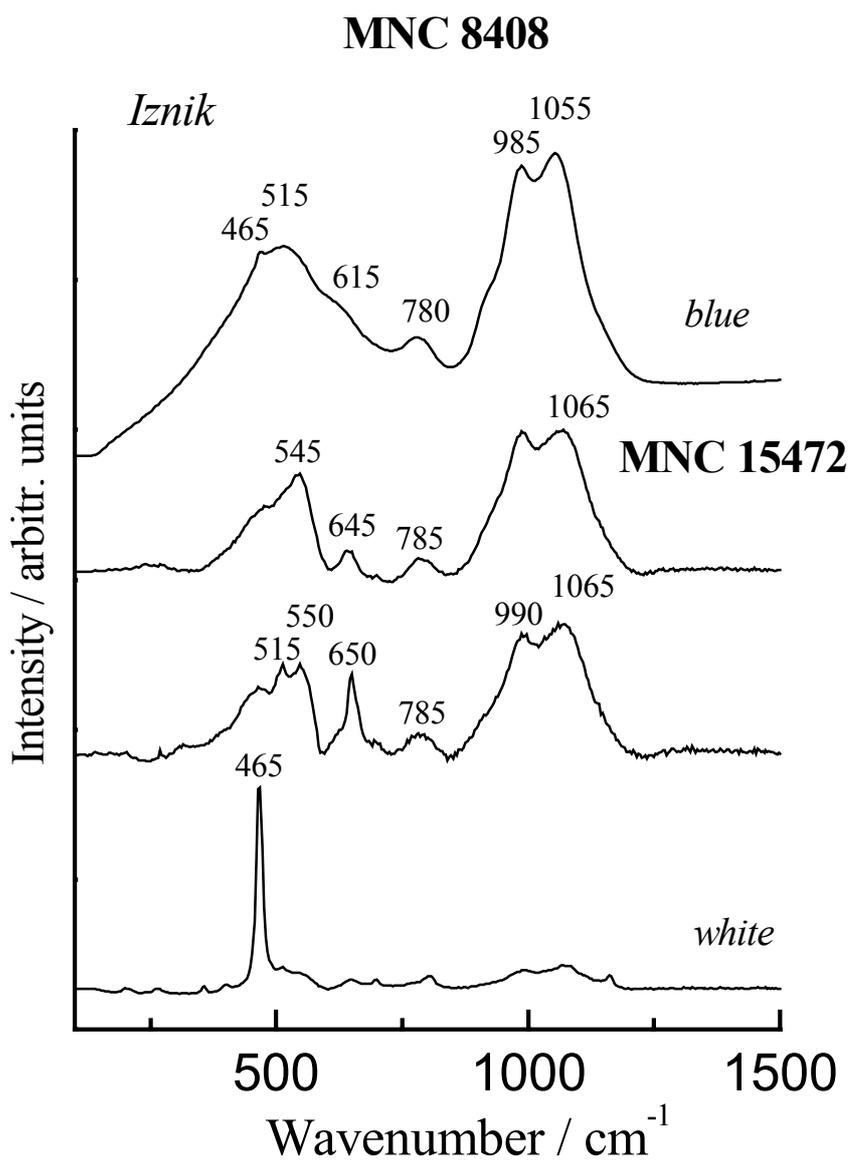

Fig.1

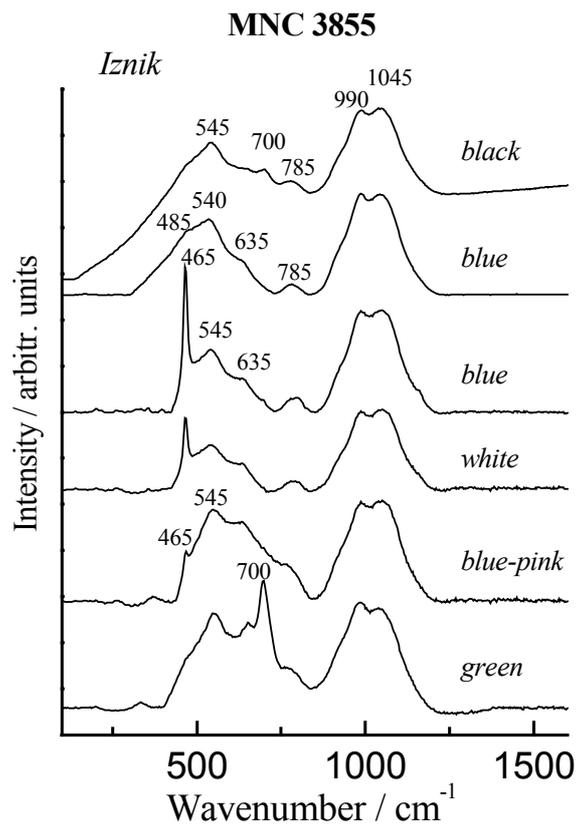

Fig. 2

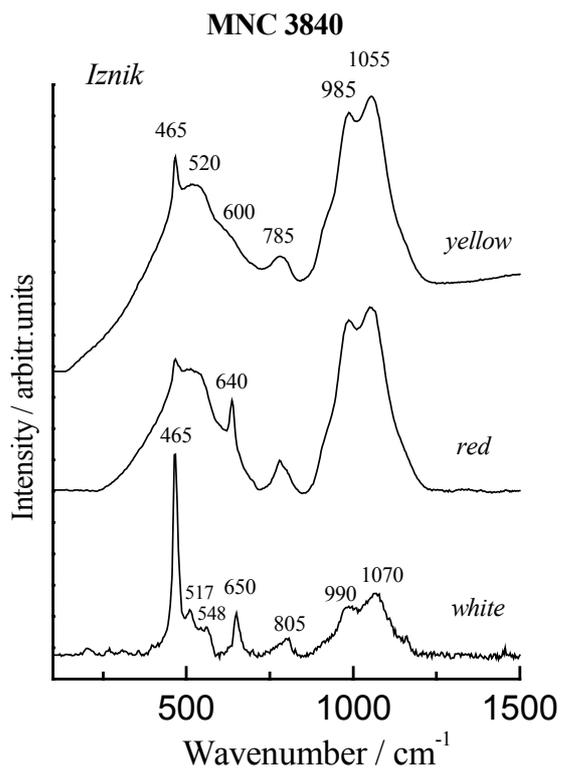
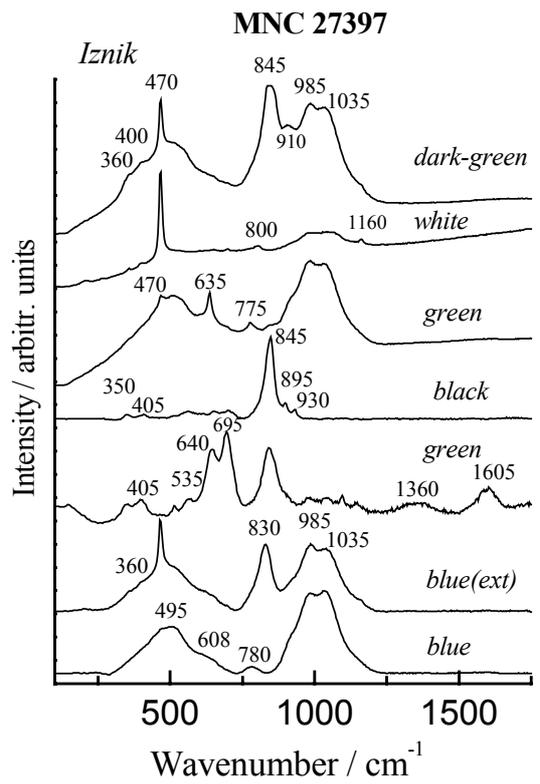

Fig. 3

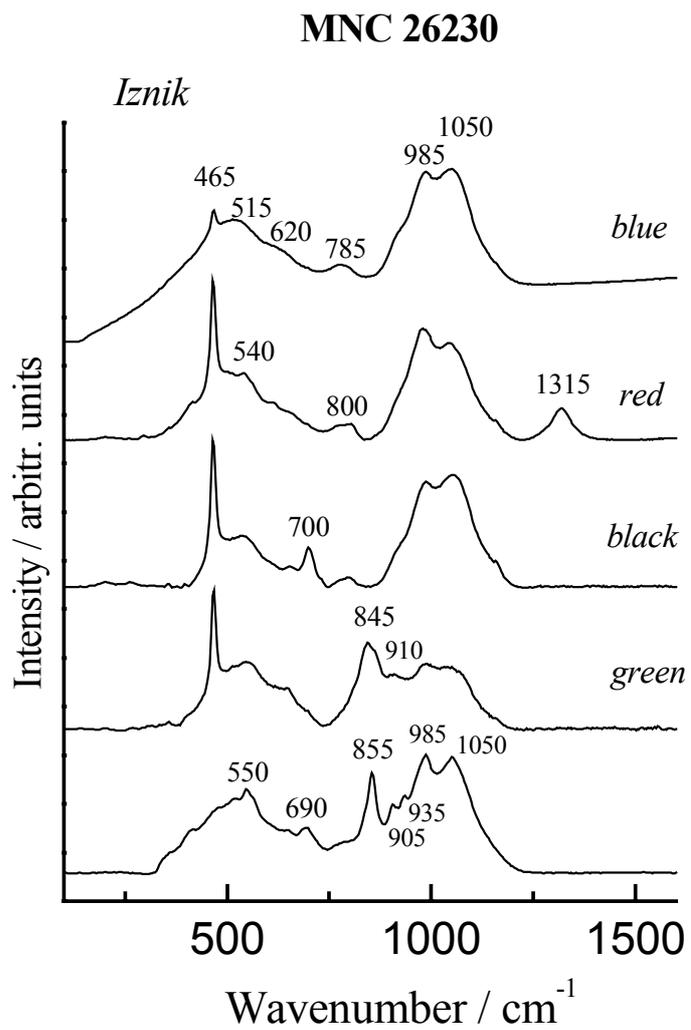

Fig. 4

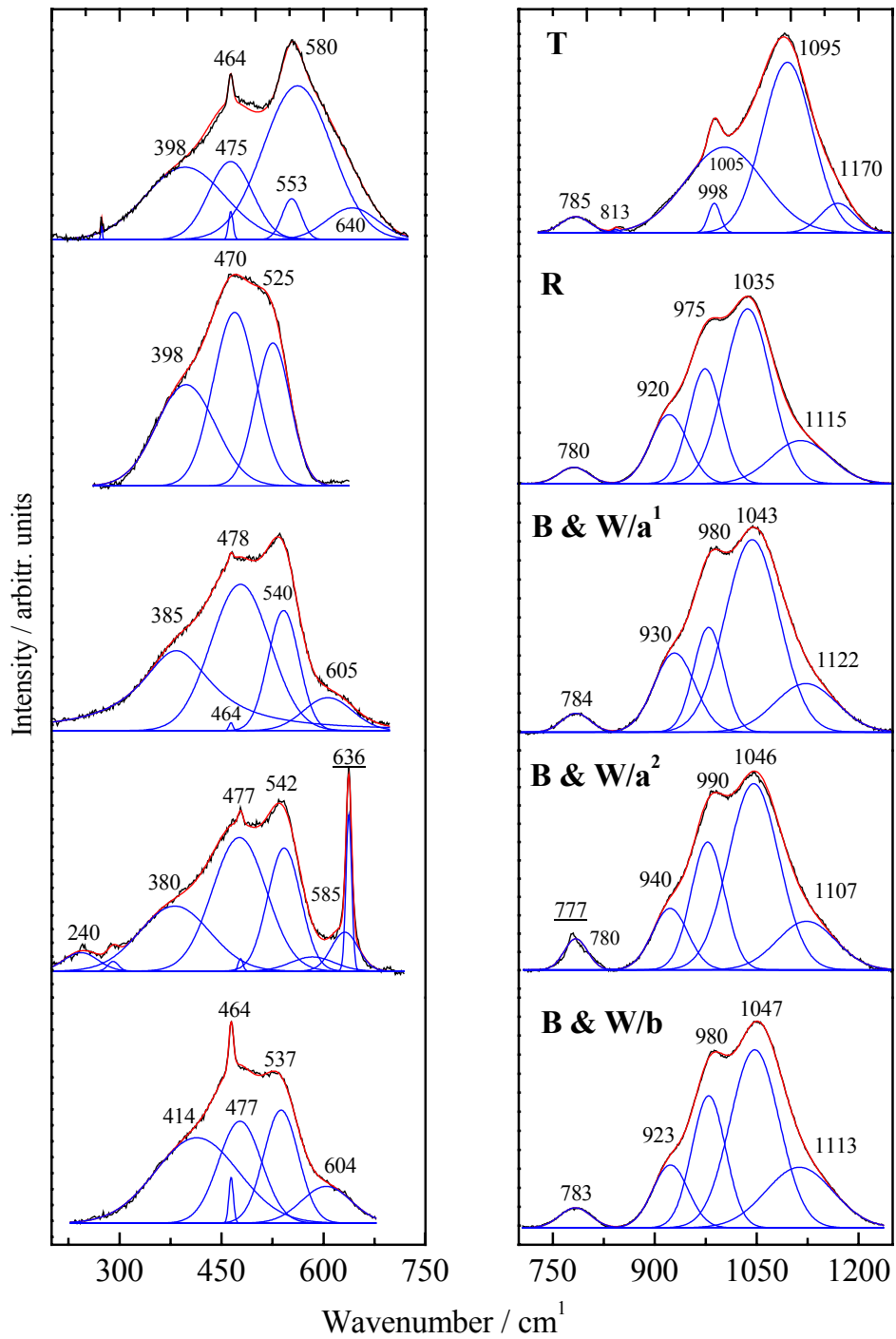

Fig. 5

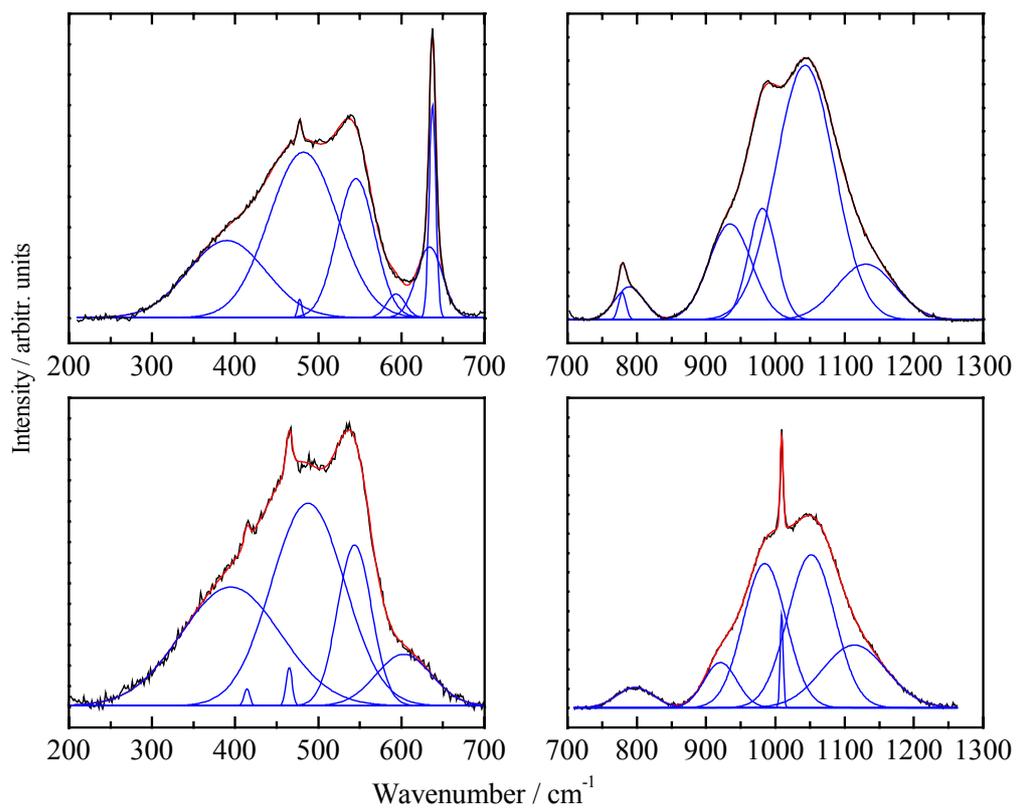

Fig. 6

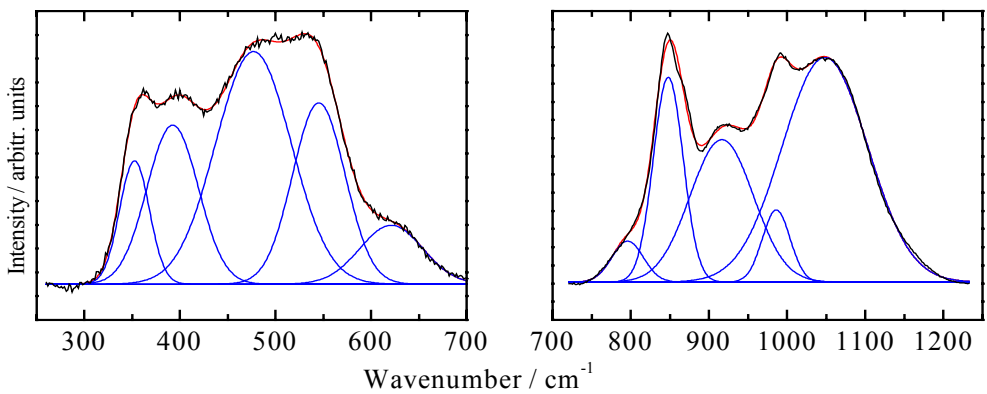

Fig. 7

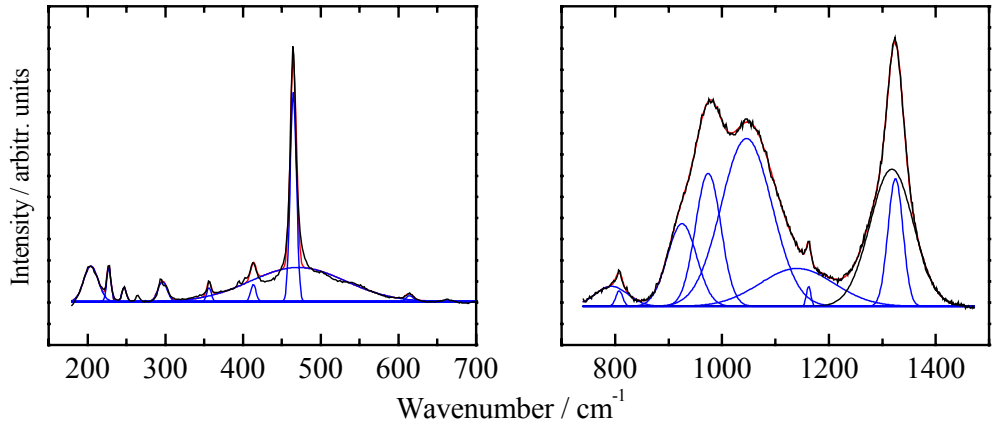

Fig. 8

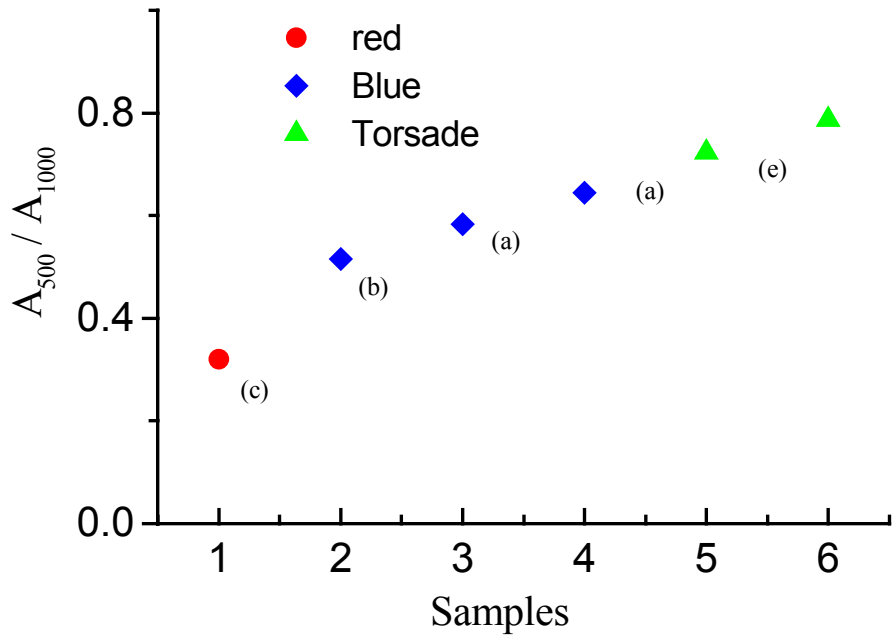

Fig. 9

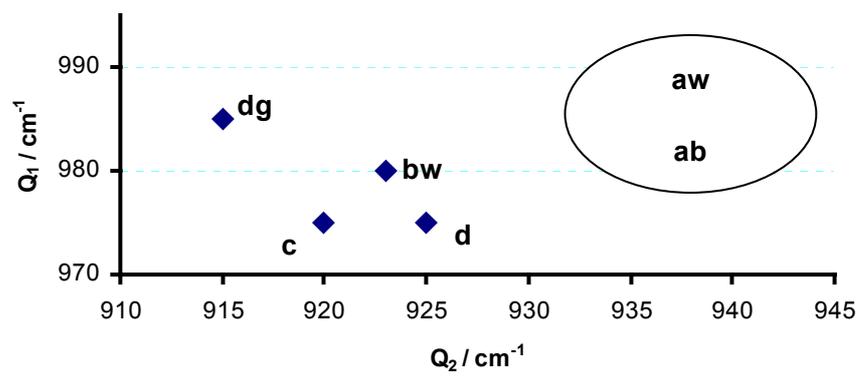

Fig. 10